\documentclass[12pt]{article}

\usepackage{graphicx}
\usepackage{amssymb}

\newcommand{\rf}[1]{(\ref{#1})}
\newcommand{\bea}{\begin{eqnarray}}
\newcommand{\eea}{\end{eqnarray}}

\newcommand{\e}{\mbox{e}}
\renewcommand{\d}{\mbox{d}}

\newcommand{\G}{\Gamma}

\renewcommand{\L}{\Lambda}

\newcommand{\n}{\nu}
\newcommand{\m}{\mu}

\renewcommand{\th}{\theta}
\newcommand{\Th}{\Theta}
\newcommand{\ep}{\varepsilon}

\newcommand{\del}{\delta}
\newcommand{\Del}{\Delta}

\newcommand{\oh}{\frac{1}{2}}
\newcommand{\oq}{\frac{1}{4}}
\newcommand{\dg}{\dagger}

\newcommand{\tr}{\mathrm{Tr}\,}

\newcommand{\prt}{\partial}
\newcommand{\mi}{\!-\!}
\newcommand{\equ}{\!=\!}

\newcommand{\hx}{{\hat{x}}}
\newcommand{\hy}{{\hat{y}}}
\newcommand{\hph}{{\hat{\phi}}}
\newcommand{\hDel}{{\hat{\Del}}}
\newcommand{\hD}{{\hat{D}}}
\newcommand{\hprt}{{\hat{\prt}}}

\newcommand{\hZ}{{\hat{Z}}}

\def\void{}
\def\labelmark{}

\newenvironment{formula}[1]{\def\labelname{#1}
\ifx\void\labelname\def\junk{\begin{displaymath}}
\else\def\junk{\begin{equation}\label{\labelname}}\fi\junk}%
{\ifx\void\labelname\def\junk{\end{displaymath}}
\else\def\junk{\end{equation}}\fi\junk\labelmark\def\labelname{}}

{\ifx\void\labelname\def\junk{\end{array}\end{displaymath}}
\else\def\junk{\end{array}\right.\end{equation}}
\fi\junk\labelmark\def\labelname{}\def\junk{}
}

\newcommand{\beq}{\begin{formula}}
\newcommand{\eeq}{\end{formula}}
\newcommand{\beqv}{\begin{formula}{}}

\begin{document}

\hfill SU-4252-767\\ 

\hfill  August 2002

\begin{center}
\vspace{24pt}
{ \large \bf  Stripes from (noncommutative) stars}

\vspace{30pt}

{\sl J. Ambj\o rn}$\,^{a,}$ and {\sl S. Catterall}$\,^{b}$

\vspace{24pt}

$^a$~The Niels Bohr Institute, \\
Blegdamsvej 17, \\
DK-2100 Copenhagen \O , Denmark\\
{\it email: ambjorn@nbi.dk}

\vspace{10pt}
$^b$~Department of Physics, \\
Syracuse University,\\
Syracuse NY 13244, USA\\
{\it email: smc@physics.syr.edu}

\vspace{48pt}

\end{center}


\begin{center}
{\bf Abstract}
\end{center}

We show that lattice regularization of noncommutative 
field theories can be used to study non-perturbative 
vacuum phases. Specifically we provide evidence for the existence of
a {\it striped} phase in two-dimensional noncommutative scalar
field theory.

\vspace{12pt}
\noindent


\newpage

\subsection*{Introduction}\label{intro}

The mixing of UV and IR physics in noncommutative field
theories is a intriguing phenomenon. One of the more spectacular 
manifestations of this mixing is the appearance of a so-called 
striped phase which breaks translational invariance in noncommutative 
$\phi^4$ theory. The prediction of this phase was based on a self-consistent 
Hartree treatment of one-loop diagrams \cite{gubser} or 
a one-loop renormalization group approach \cite{cw}. The purpose of the 
following note is to test the predictions using the non-perturbative 
framework of lattice computer simulations and at the same time test the 
viability of studying noncommutative field theories by numerical 
means. 
Here we consider the simplest possible such theory: a two-dimensional
$\phi^4$ theory. Two-dimensional chiral and Yang-Mills theories have
also been studied recently \cite{chiral,yang}.

\subsection*{The model}\label{model}

Our starting point is noncommutative scalar field theory 
with the action
\beq{1}
S = \int \d^2 x \left[ \oh (\partial_{\mu}\phi)^2 + \oh m^2 \phi^2 
+\frac{g^2}{4} \phi^{*4}\right],
\eeq
where $\phi^{*4} = \phi * \phi * \phi * \phi$ and where the 
star product is defined by 
\bea
(\phi_1 * \phi_2)(x) &=& 
\e^{ \frac{i}{2} \th^{\m\n} \prt_{x_\m} \prt_{y_\n}} \phi_1(x)\phi_2(y)
\Big|_{x=y}\\
&=&  \frac{1}{\pi^2 |\det \th|} \int\int \d^2y\d^2z \,\phi_1(y)\phi_2(z)\, 
\e^{2i \th^{-1}_{\m\n}(x-y)_\m(x-z)_\n}. \label{2a}
\eea
In particular we have 
\beq{3}
[x_\m,x_\n]_* = x_\m * x_\n -x_\n * x_\m = i \th_{\m\n} = i \th \ep_{\m\n}.
\eeq

It is known how to map formally the noncommutative theory into a 
matrix theory by the Weyl map. Let $\hx^\m$ and $\hy^\n$ be Hermitean 
operators satisfying $[\hx_\m,\hy_\n]= i \th_{\m\n}$, which 
is the operator analogy of \rf{3}. One can define the Weyl map by 
\beq{5}
\hph = \int \d^2x \,\phi(x)\ \,\hDel (x),~~~~~
\hDel = \int \frac{\d^2 k}{(2\pi)^2} \; \e^{ik_\m \hx_\m}\e^{-ik_\n x_\n}.
\eeq  
and in this way the action \rf{1} can be represented (formally, with 
a suitable normalization of the trace and suitable assumptions concerning
the fall off of the functions $\phi(x)$ at infinity) as 
\beq{7}
S = \tr \left( \oh \left[ \hprt_\m,\hph\right]^2+\oh m^2 \hph^2 + 
\frac{g^2}{4} \hph^4\right),
\eeq
where $\hprt_\m$ is an anti-hermitian derivative which together 
with $\hx_\m$ satisfy
\beq{8}
[\hprt_\m,\hx_\n]= \del_{\m\n},~~[\hprt_m,\hprt_\n]= ic_{\m\n},
~~~~[\hx_\m,\hx_\n]=i\th_{\m\n}.
\eeq

The operators (or matrices) in \rf{8} are all infinite dimensional.
However, the associated Heisenberg algebras of 
exponentials of $\hx_\m$ (and $\hprt_\m$) 
allow finite dimensional representations,
as already noted by Weyl. This is the key to formulating the 
noncommutative theory on a finite periodic $N\times N$ lattice
with lattice spacing $a$ 
(see \cite{amns3} for a more general formulation). The algebra
\rf{8} is replaced by
\beq{9}
\hZ_\m \hZ_\n = \e^{-2\pi i \Th \ep_{\m\n}} \hZ_\n\hZ_\m,~~~
\hD_\m \hZ_\n \hD_\m = \e^{2\pi i\delta_{\mu\nu}/N} \hZ_\n,
\eeq
where 
\beq{10}
\hZ_\m = \e^{2\pi i \hx_\m/a N},~~~\hD_\m = \e^{a \hprt_\m},~~~
\Th = \frac{2\pi \th}{a^2N^2}.
\eeq
Periodicity on the lattice fixes the quantization of momentum $p_\m$ and $\th$
\beq{11}
p_\m a = k_\m = \frac{2\pi m_\m}{N},~~~~\th = \frac{n a^2 N}{\pi}~~~
\left(\Th = \frac{2n}{N}\right),
\eeq
where $m_\m \in 0,\ldots,N\mi 1$, and $n$ is an integer.  
In the following we choose $n \equ 1$ for simplicity.

A simple realization of the above scenario is obtained by 
choosing $\hZ_\m$, $\m \equ 1,2$ to be the standard $N\times N$ shift and
clock matrices $(\hZ_1)_{jk}= \del_{j+1,k}$, 
$(\hZ_2)_{jk}= (\e^{4\pi i/N})^{j-1}\del_{jk}$, $\hD_1 = (\hZ_2^\dg)^{(N+1)/2}$,
$\hD_2 = (\hZ_1)^{(N+1)/2}$, the lattice size $N$ being odd. 

In this way the lattice version of \rf{2a} becomes
\beq{12}
\phi_1(x) * \phi_2(x) = \frac{1}{N^2}
\sum_{y,z} \phi_1(y)\phi_2(z)\;\e^{2i\th^{-1}_{\m\n}(x-y)_\m(x-z)_\n},
\eeq
and the operator $\hDel$ becomes  an $N\times N$ matrix 
\beq{13}
\hDel(x) = 
\sum_{m_1,m_2 = 1}^{N} \left(\hZ_1^{m_1}\hZ_2^{m_2} \;
\e^{-\pi i \Th m_1m_2}\right) \e^{-2\pi i m_\m x_\m/aN}.
\eeq
It generates an explicit map from the lattice field $\phi(x)$ defined 
at the $N\times N $ lattice points $x$ to that $N\times N$ matrix $\hph$
defined by \rf{5},
\beq{14}
\hph = \frac{1}{N^2}\sum_x \phi(x) \hDel(x),~~~~\phi(x) = \frac{1}{N}\tr \hph \hDel (x)
\eeq
such that the action is given by \rf{7} with $N\times N$ matrices and 
the partition function is 
\bea\label{15}
Z(R,G) &=& \int \d \hph \; \e^{-S[\hph]},\\
S[\hph] &=& N \tr \left[ \oh\sum_\m\left(\hD_\m\hph \hD_\m^\dg\mi \hph\right)^2
+  R \hph^2 + G\hph^4\right]\label{16}.  
\eea
The dimensionless lattice parameters are 
\beq{15a}
R= \oh m^2 a^2,~~~~~G = \oq g^2 a^2.
\eeq
The ordinary (commutative) $\phi^4$ theory on the lattice has two 
different continuum limits when the space-time volume goes to 
infinity, one ``trivial'' associated with  the fine tuning the 
approach to $R \equ 0$ and $G \equ 0$. It corresponds to the 
standard $\phi^4$ two-dimensional field theory where only 
normal ordering is needed to renormalize the theory. The other
limit is associated with the phase transition between a phase in
which the symmetry $\phi\to -\phi$ is broken and a symmetric
phase. 
It occurs for finite negative values
of $R$ and finite positive values of $G$ and the continuum limit 
can be identified with the continuum limit of the Ising model,
i.e.\ with a $c \equ 1/2$ conformal field theory.
 
In the noncommutative case the continuum limit of course requires 
an infinite lattice, i.e. we have to take $N$ to infinity. 
In addition we  are interested in a limit in which the lattice 
spacing can be viewed as scaling to zero while 
the dimensionful parameter $\th$ stays finite. Two possible
scenarios appear possible:

If non-commutativity can be considered as a perturbation 
imposed on a commutative theory then a kind of ``double scaling 
limit'' of the matrix model  \rf{16} suggests itself:
\beq{17}
N \to \infty,~~~~~a^2 =\frac{\pi\th}{N},~~~~~ a^2N^2 = \pi \th N. 
\eeq
Thus the physical space-time diverges like $N$ \footnote{For 
other limits which allow for a finite physical space-time
see \cite{amns1,amns2,amns3}.} and we can write the matrix model 
action \rf{16} as 
\beq{17a}
S[\hph] = N \tr \left[ \oh\sum_\m\left(\hD_\m\hph \hD_\m^\dg\mi \hph\right)^2
+ \frac{\pi m^2 \th}{2N} \hph^2 + \frac{\pi g^2 \th}{4N}\hph^4\right]. 
\eeq
The noncommutative parameter $\th$ only appears in the dimensionless
combinations $m^2 \th$ and $g^2\th$ and implicitly in factors of 
$N$ appearing in \rf{17a}. This type of continuum limit
seems to be most naturally associated with the trivial
gaussian fixed point of the commutative theory.
Note that this ``double scaling'' limit 
$N \to \infty$ is quite different from the usual planar limit of 
matrix models and thus also quite different
from the double scaling limit of the 2d matrix models studied in the 
context of 2d quantum gravity.

If, on the other hand, non-commutativity alters 
the commutative theory profoundly one 
should follow the general strategy of non-perturbative lattice 
studies in order to identify a continuum limit: identify suitable 
physical observables usually related to a divergent correlation length
of the lattice field. Fine-tuning the coupling constants such that 
the observables are constant when expressed in terms of the 
``physical'' correlation length $a \cdot \xi$, where $\xi$ is 
the correlation length in lattice units, will then lead to a 
relation between the coupling constants and the lattice spacing $a$
eg. $a\sim\left(R-R_c\right)^\nu$
and allow us to identify the (potentially) non-perturbative renormalization
of the coupling constants required to obtain a continuum theory.
We can then potentially investigate another
``double scaling'' limit in which $N\left(R-R_c\right)^{2\nu}\sim \th$ is
held constant as $N\to\infty$ and $R\to R_c$. Such a phase transition
would generalize the usual Ising transition of the commutative
theory to the noncommutative case.

The first scenario has the attractive feature that we know ahead of
time the value of the non-commutativity parameter $\theta$ and the
dependence of lattice spacing on the matrix size $N$. We will show
numerical evidence which supports scaling according to
this scenario although we stress
that the existence of such a double scaling limit 
does not rule out the possible existence of additional continuum
limits based on the second scenario.

Susceptibilities (i.e. correlation functions 
integrated over all space-time points)
have been used in conventional lattice field theory to identify 
the couplings associated with a divergent correlation length, and 
the critical indices associated with the susceptibilities lead
to a classification of the continuum Euclidean field theories.
Susceptibilities are suitable observables also in the noncommutative 
case. In fact one has 
\beq{17x}
\int \d^d x \; \phi(x) * \phi(x) = \int \d ^d x \; \phi(x) \phi(x),
\eeq
so they are maximally close to the commutative observables. Yet
they may contain additional interesting information about the 
infrared behavior of the noncommutative propagators. In the computer 
simulations to be reported below
we will indeed use (generalized) susceptibilities to 
identify potential phase transitions. 
\begin{eqnarray*}
\chi_p&=&\left<\int d^2 x e^{ip.x} \phi(x) \int d^2 y e^{-ip.y} \phi(y)\right>\\
      &=&\left<\overline\phi(p)\overline\phi(-p)\right>    
\end{eqnarray*} 
Physically, they are just the (mod square) of the Fourier component of the
noncommutative field.

\subsection*{Stripes}

The possibility of having a ground state which breaks translational
invariance was first observed in \cite{gubser}. The quadratic part 
of the effective action,
obtained from \rf{1} by a  one-loop self-consistent Hartree calculation,
is   
\beq{18}
\G^{(2)}(p) = p^2+m_R^2 + A g^2 \ln \L_p^2,~~~~~
\L_p^2= \frac{1}{\th^2p^2 + 1/\L^2}.
\eeq	
$\L$ denotes a UV cut-off, $m_R^2$ the renormalized mass term 
and $A$ is some constant. The mixing of IR and UV modes is 
manifest in the form of $\L_p$. In addition the momentum 
dependence of $\L_p$ implies that $\G^{(2)}(p)$ can have a minimum 
for 
\beq{19}
p^2_c = Ag^2 -\frac{1}{\L^2 \th^2}.
\eeq 
For sufficiently large coupling this analysis would predict
that modes with momentum $p=p_c$ would be the first to become unstable as
the mass parameter $m_R$ were tuned to zero. Thus any phase transition would
then be driven by condensation of a non-trivial momentum component
of the field. This is to be contrasted with the usual Ising 
transition in which the $p=0$ component of the field condenses to yield
the broken symmetry phase.
Using \rf{15a} and \rf{17} this result can be written in lattice
variables
\beq{20}
\left(\frac{2\pi n_c}{N}\right)^2 = 
A \,G - \frac{1}{N^2} 
\eeq 
Note that for small $G$ the matrix size $N$ has to be large  ($N > 2\pi/\sqrt{A \,G}$)
in order for \rf{20} to have a solution. Conversely, we might
expect an exotic vacuum state to exist for {\it any} coupling in
the continuum limit $N\to\infty$.

In two dimensions it is usually difficult to have  a 
spontaneous breaking of a continuous symmetry 
like translation invariance. The fluctuations associated 
with the corresponding massless Goldstone boson will 
spoil any long range order. However, the arguments leading 
to this conclusion are strictly speaking not valid for 
non-local field theories and there exist explicit examples
of non-local field theories in two dimensions which 
exhibit spontaneous symmetry breaking\footnote{One example is the 
so-called crumpling phase transition in crystalline membranes.}
Since noncommutative field theories are non-local and 
show a mixing of IR and UV properties one cannot rule out
that the simple one-loop arguments given above in favor of
a non-translational invariant ground state could still be valid.
In fact the renormalization group treatment in \cite{cw}
predicts this ground state even in two dimensions. 

In the following we will look for signals indicating the appearance of 
a non-trivial ground state in the lattice system \rf{16}.   
  
\subsection*{The computer simulation}

In these first explorative simulations we have used a simple 
metropolis algorithm for stimulating the matrix system \rf{15}-\rf{16}.
Specifically, the matrix field is updated by the addition of
a random hermitian noise matrix $\eta$
\begin{equation}
\phi\to \phi+\epsilon\eta
\end{equation}
We tune $\epsilon$ with $N$ so as to achieve an acceptance rate close
to 50\%. Most of our runs employed $5\times 10^6$ such matrix updates
for a given set of values of the parameters $m$ and $g$. 
If we only consider the quadratic part of \rf{16} the system is {\it
equivalent} to an ordinary gaussian lattice field theory via 
the map \rf{14} and this has been a useful check of the computer code. 

Let us now turn to the actual simulation with the noncommutative $\phi^{*4}$
term. We have looked in the ($m$,$g$) coupling constant space for signals
of a nontrivial vacuum.
We have not attempted to map the entire
phase diagram but concentrated instead on the slice
$\theta=1.0$ and $g^2/4=1.0$ with varying physical
mass parameter $m$.
\begin{figure}[htb]
\begin{center}
\includegraphics[width=13cm]{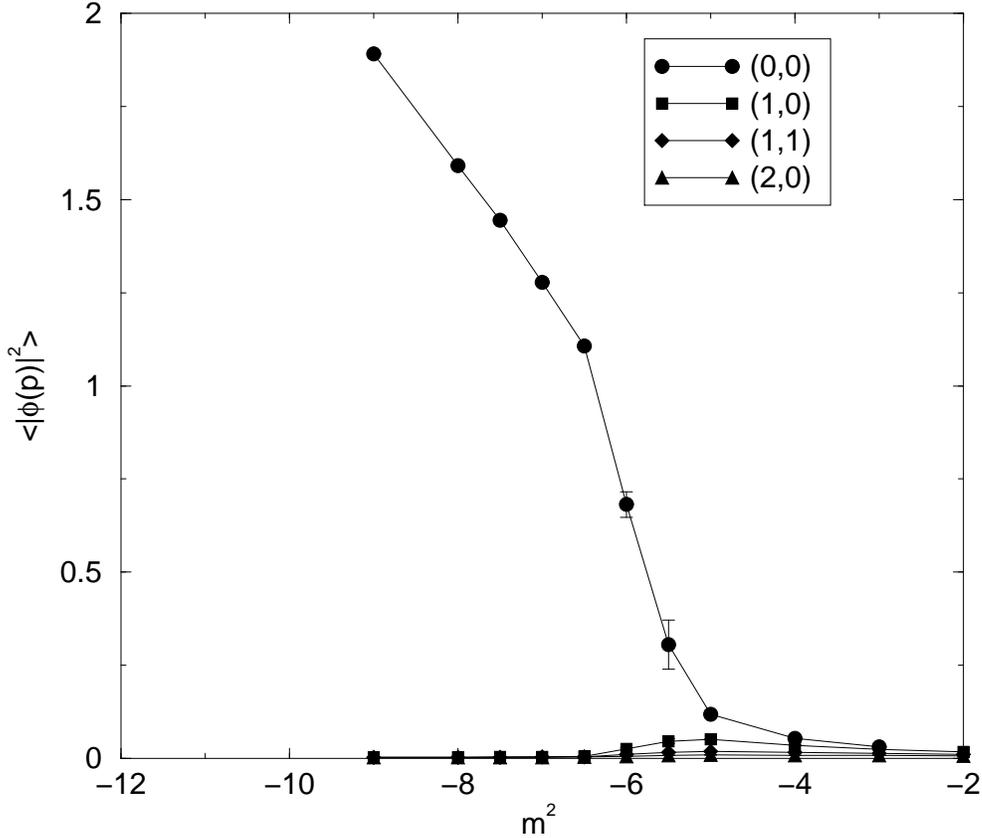}
\end{center}
\caption{the average of the (mod square) of the
Fourier amplitude for 4 lattice modes vs mass parameter $m^2$ (for
lattice size $N=9$). The 4 modes
correspond to momentum states \rf{21} where
$(m,n)$ is $(0,0),(1,0),(1,1)$ and $(2,0)$.}
\label{fig1}
\end{figure}
In Fig.\ \rf{fig1} we show the generalized susceptibility  
$\chi_k$ corresponding
to the (mod squared) Fourier component of the field $\phi (x)$ on
an $N=9$ lattice 
for various values of the lattice momentum 
\beq{21}
(k_1,k_2) = \left(\frac{2\pi m}{N},\frac{2\pi n}{N}\right),~~~~(p= k/a).
\eeq
\begin{figure}[htb]
\begin{center}
\includegraphics[width=13cm]{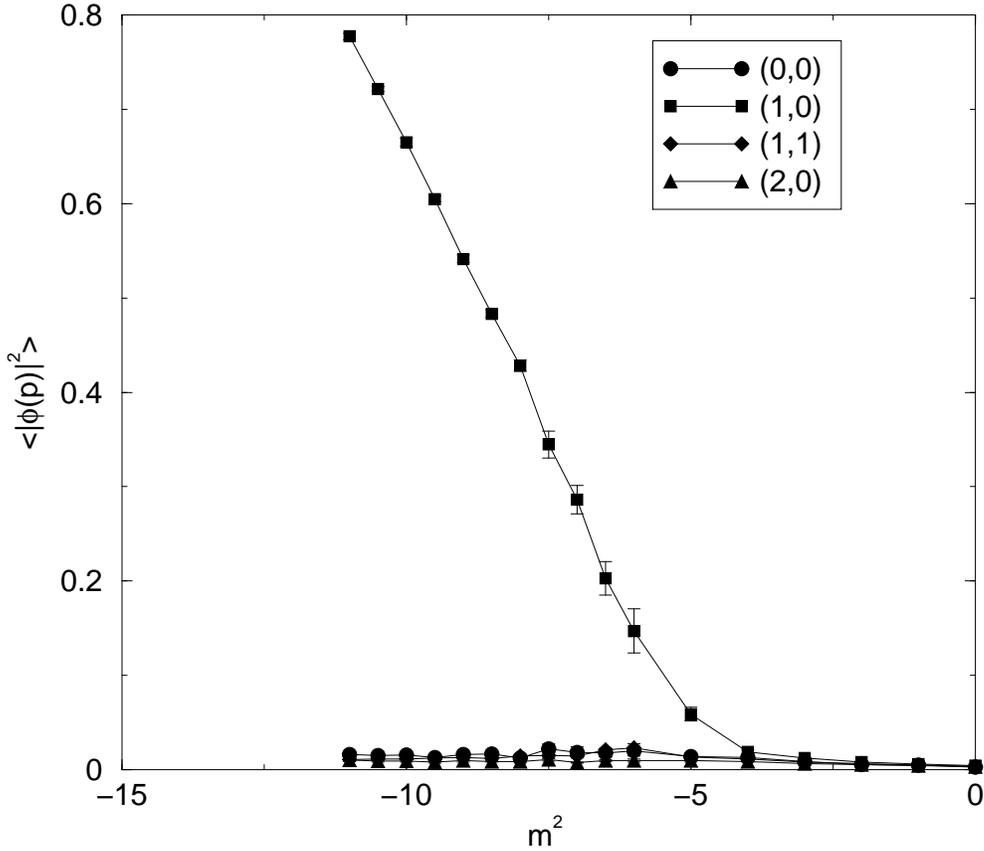}
\end{center}
\caption{\label{fig2}the average of the (mod square) of the
Fourier amplitude for 4 lattice modes vs mass parameter $m^2$ (for
lattice size $N=23$). The 4 modes
correspond to momentum states \rf{21} where 
$(m,n)$ is $(0,0),(1,0),(1,1)$ and $(2,0)$.}
\end{figure}
One sees a clear condensation of the usual $p=(0,0)$ mode for values
of $m^2 < -6$. Contrast this with the Fig.\ \rf{fig2} which shows
the same set of (unsubtracted) susceptibilities for $N=23$.  
Now for $m^2 <-6$ we see that the dominant susceptibility
corresponds to the lattice
$(1,0)$ mode -- the vacuum in this
region of the parameter space shows a spatially varying 
order parameter as predicted by the one loop calculation - a striped
phase. Since the system still possesses a symmetry under
exchange of axes we find that the condensation takes place along
either the 1-axis or 2-axis randomly as the mass is varied. Thus,
in our plots we show the average
$<\vert\overline{\phi}(n,m)\vert^2+\vert\overline{\phi}(m,n)\vert^2>$.
\begin{figure}[htb]
\begin{center}
\includegraphics[width=13cm]{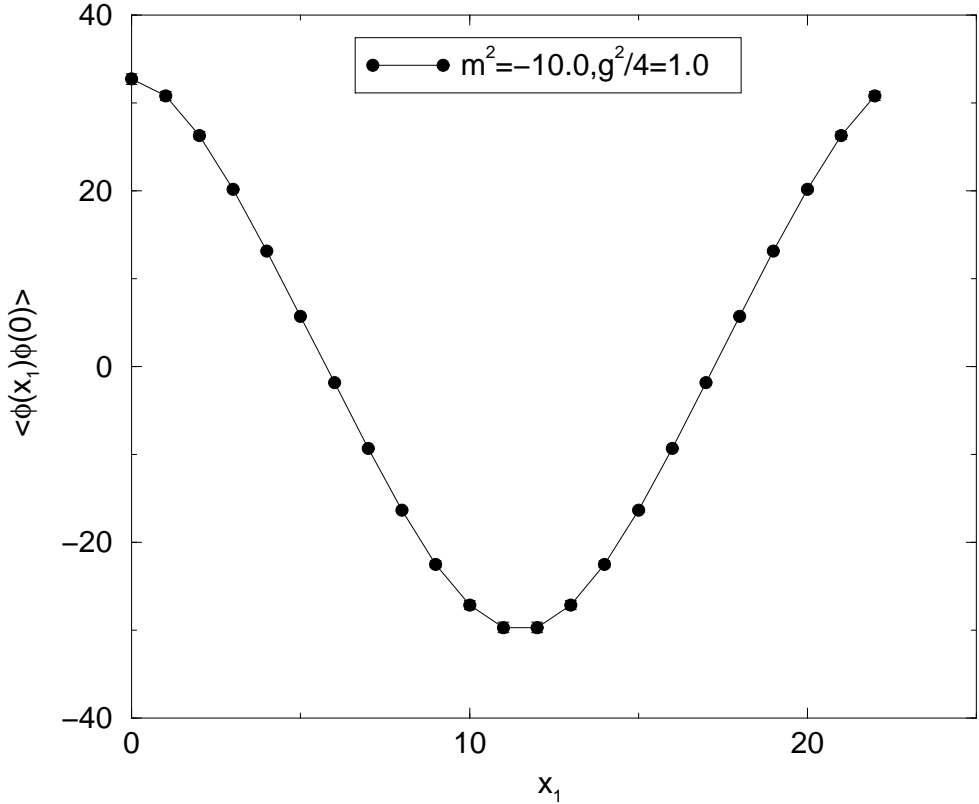}
\end{center}
\caption{\label{fig3}$\langle\phi(x_1)\phi(0)\rangle$ vs. $x_1$ for 
$m^2=-10.0$ and $N=23$}
\end{figure}
\begin{figure}[htb]
\begin{center}
\includegraphics[width=13cm]{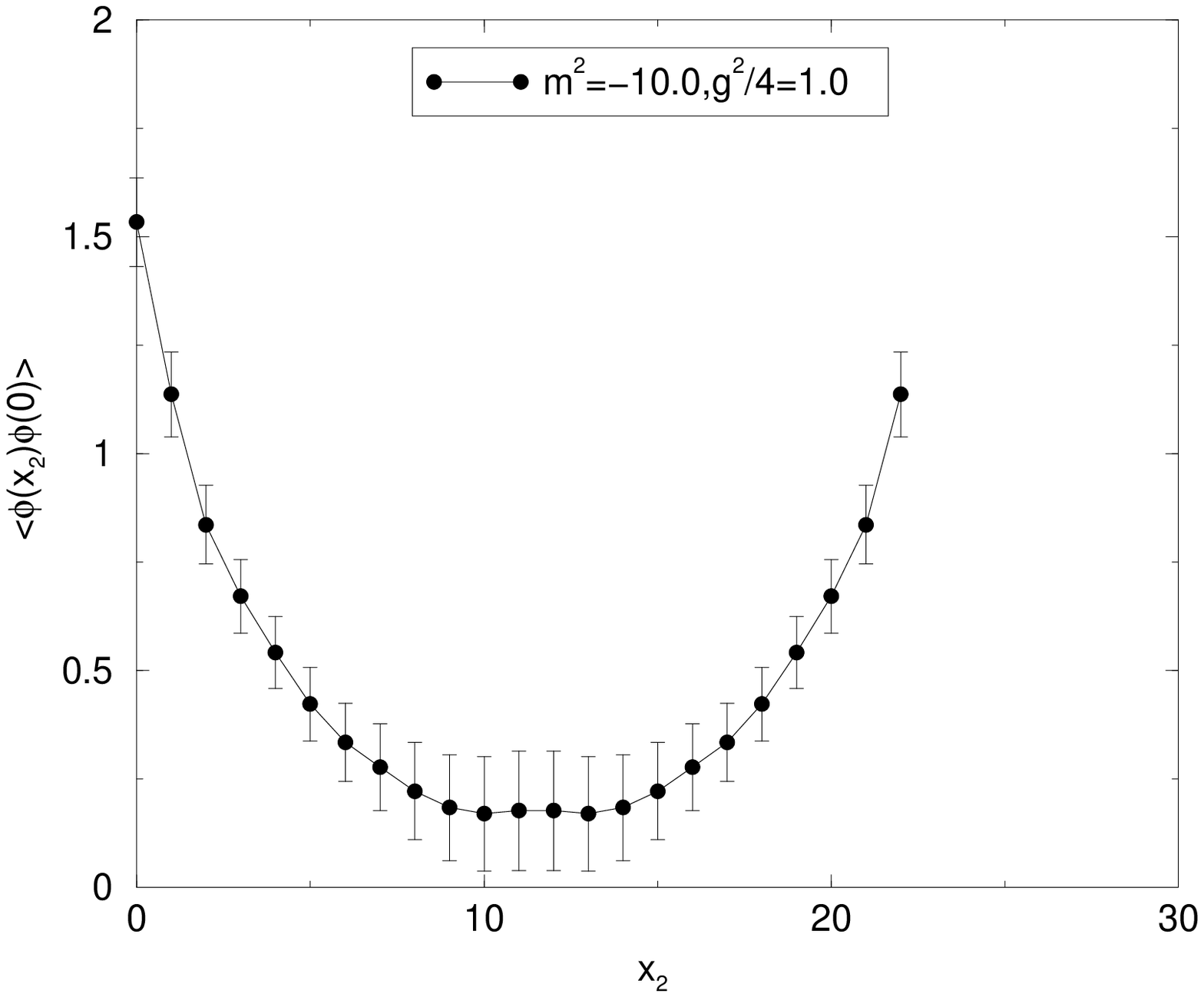}
\end{center}
\caption{\label{fig4}$\langle\phi(x_2)\phi(0)\rangle$ vs. $x_2$ for $m^2=-10.0$ 
and $N=23$}
\end{figure}
Another way to illustrate graphically the existence of this phase is
by considering the transverse-averaged correlator in position
space. Fig.\ \rf{fig3} shows a plot of $<\phi(x_1)\phi(0)>$ for
$m^2=-10.0$. Notice that the correlator shows one complete oscillation
within a lattice length. Since we are deep within the striped
phase this correlator will essentially measure the behavior of the order
parameter directly. Along the second direction the correlator is
positive definite as shown in Fig.\ \rf{fig4}.
\begin{figure}[htb]
\begin{center}
\includegraphics[width=13cm]{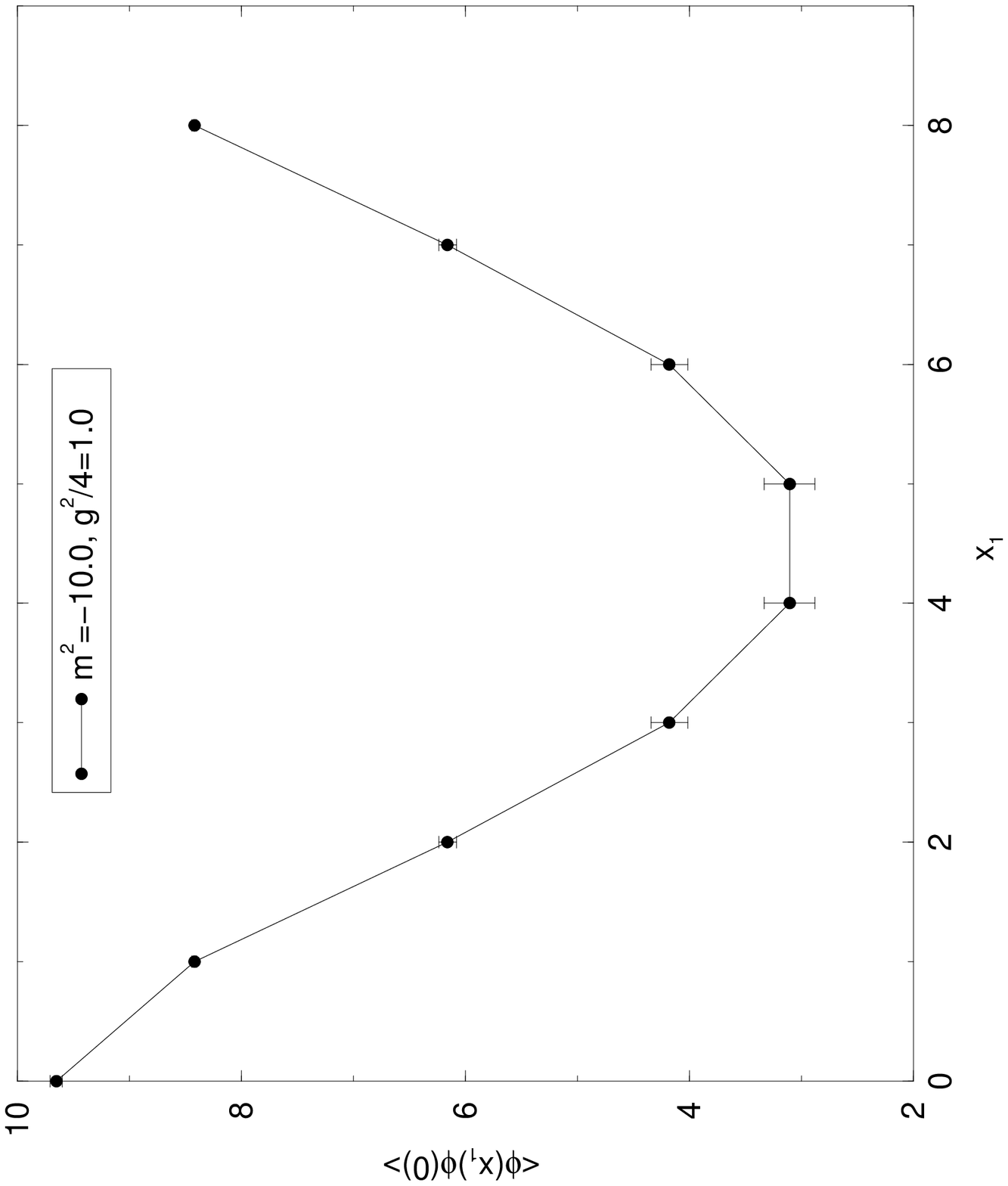}
\end{center}
\caption{\label{fig5}$\langle\phi(x_1)\phi(0)\rangle$ vs. $x_1$ for $m^2=-10.0$
and $N=9$}
\end{figure}
This behavior is what we would expect of a striped phase. For the smaller
lattice $N=9$ we do not see any asymmetric behavior in the form
of this position space correlator in the two directions, and indeed,
as Fig.\ \rf{fig5} indicates, the correlator is of conventional
(positive definite) form over the entire lattice.
This is consistent with the idea that for small $N$, there
are no solutions to \rf{18} and the minimum of
the quadratic effective action is still found for mode $(0,0)$. 
\begin{figure}[htb]
\begin{center}
\includegraphics[width=13cm]{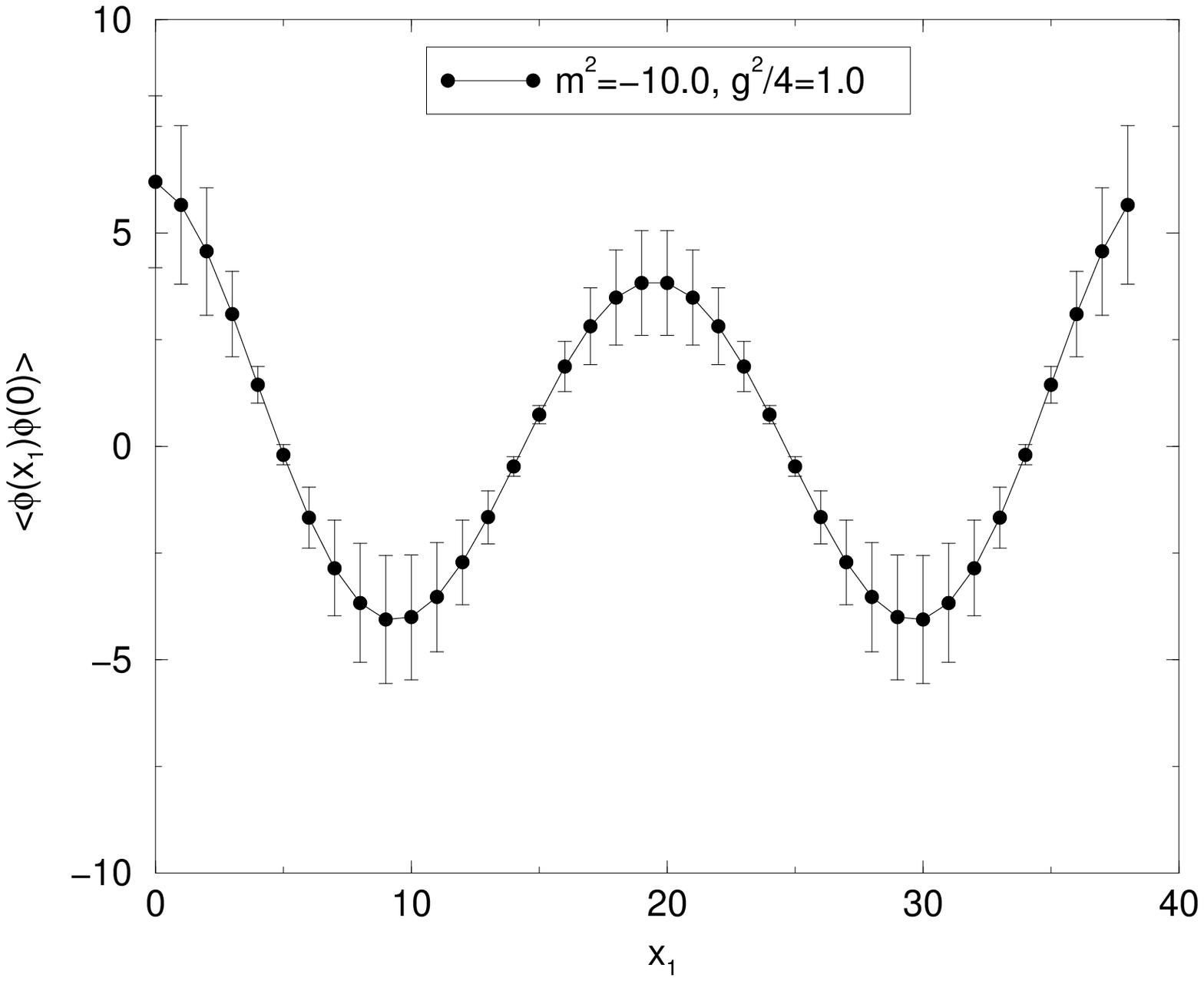}
\end{center}
\caption{\label{fig6}$\langle\phi(x_1)\phi(0)\rangle$ vs $x_1$ for $m^2=-10.0$ and
$N=39$}
\end{figure}
\begin{figure}[htb]
\begin{center}
\includegraphics[width=13cm]{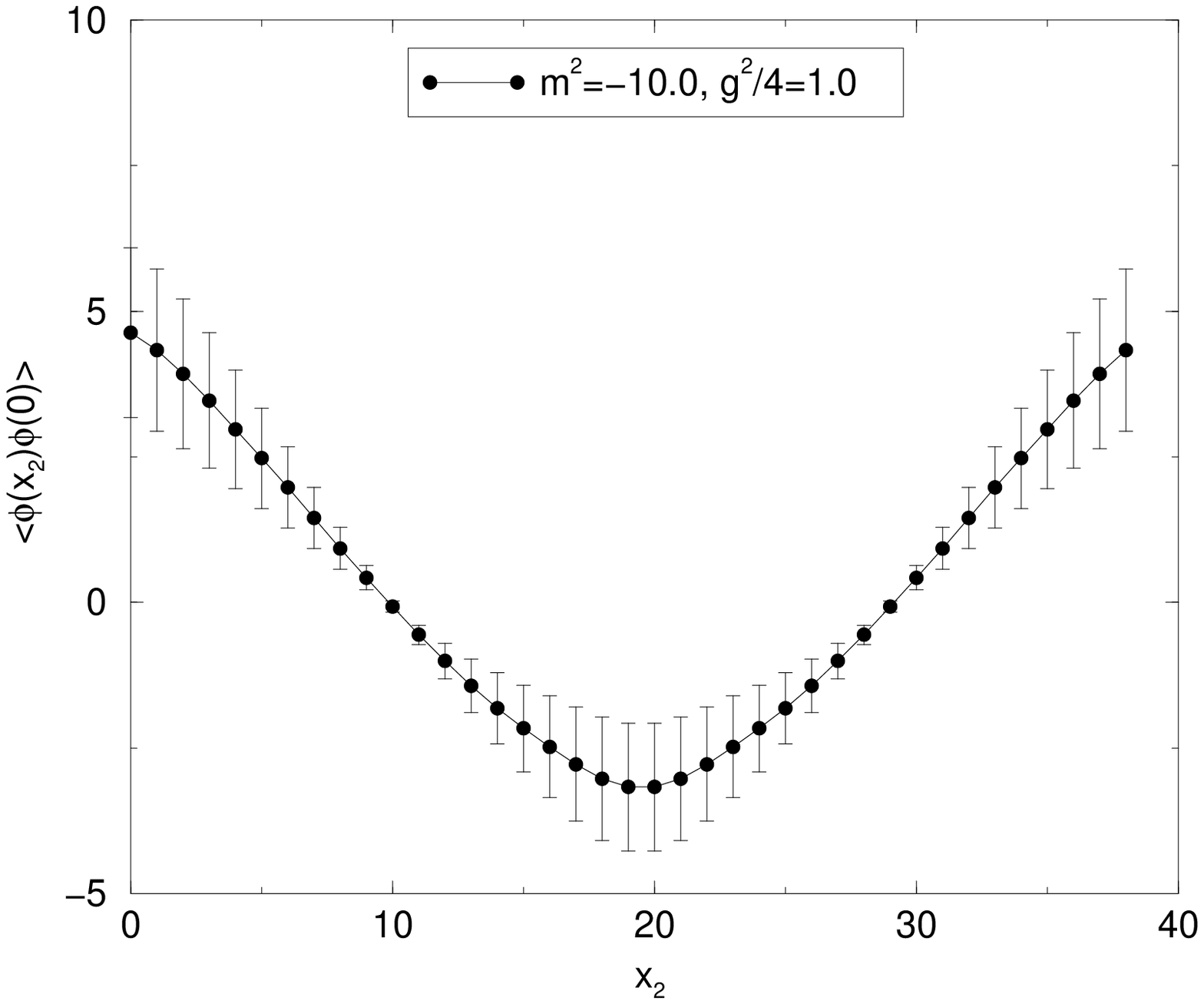}
\end{center}
\caption{\label{fig7}$\langle\phi(x_2)\phi(0)\rangle$ vs $x_2$ for $m^2=-10.0$
 and $N=39$}
\end{figure}

We have also examined the system for $N=39$. Fig.\ \rf{fig6} and
Fig.\ \rf{fig7} show the corresponding correlators 
along the two lattice axes for
$m^2=-10.0$. From
these two pictures we infer that the
condensed mode is now $(2,1)$. Thus, rather than stripes
we actually see a checkerboard-like arrangement for
the vacuum state in this case. The fact that the lattice mode
number increases with larger lattice size for fixed physical
parameters is encouraging as this is the behavior that is required if
this effect is to survive the large volume limit $N\to\infty$. Indeed,
\rf{18} indicates that the mode number $n$ should increase like 
$n\sim \sqrt{N}$ for $N\to\infty$ to ensure that the physical momentum at
which condensation occurs remains fixed.

\subsection*{Discussion}\label{discuss}

We have shown that it is indeed possible to observe 
and investigate non-trivial phenomena like the appearance 
of ``stripes'' in noncommutative field theory within 
the non-perturbative framework of lattice simulations.

We have here looked at the very simplest model, a two-dimensional
scalar field theory simply to verify that phenomena like the 
appearance of a ``striped phase'' can be observed. Our
data show the appearance of such a phase in an
appropriate double scaling limit. Notice that
the issue of possible I.R divergences in two dimensions does not
preclude such a study - the lattice regulates any such divergences
automatically and the issue returns only when the continuum limit
is taken. Our current
data do not address this issue -- much larger 
volumes must be simulated in order to
determine whether this phase truly survives the large $N$ limit. 
Nevertheless, we feel that our
simulations lend nonperturbative, albeit qualitative support
to the analysis of \cite{gubser} and \cite{cw}. 
However, it is clearly 
of more interest to do this in higher dimensional theories. The 
lattice formalism has been developed \cite{amns3} and just 
awaits use. Indeed, while this paper was in preparation we
received a preprint \cite{nishimura} in which the same model
was examined in $2+1$ dimensions with similar conclusions.

\vspace{12pt}

\subsection*{Acknowledgements} 
J.A. acknowledges the support by the
EU network on ``Discrete Random Geometry'', grant HPRN-CT-1999-00161, 
by ESF network no.82 on ``Geometry and Disorder'' and by 
``MaPhySto'', the Center of Mathematical Physics 
and Stochastics, financed by the 
National Danish Research Foundation. S.C acknowledges the support of 
DOE grant DE-FG02-85ER40237 and thanks the Niels Bohr Institute 
for hospitality during the completion of this
work.

\end{document}